\documentclass[11pt,a4paper,conference]{IEEEtran}
\usepackage{amsmath,amsthm}
\usepackage{graphicx}
\usepackage{cite}
\graphicspath{{R_plots/}}

\newtheorem{thm}{Theorem}

\def\ka{\kappa}
\def\wbeta{\tilde{\beta}}
\def\wgamma{\tilde{\gamma}}
\def\wrho{\tilde{\rho}}
\def\Rma{{\cal R}_0^{MA}}
\def\Rnet{{\cal R}_0^{NET}}
\def\R{{\cal R}_0}
\begin{document}
%
\title{Equivalence of Mass Action and Poisson Network SIR Epidemic Models}


\author{\IEEEauthorblockN{Grzegorz A. Rempa{\l}a}
\IEEEauthorblockA{Division of Biostatistics\\
The Ohio State University \\
Columbus, OH 43210 United States \\
Email: rempala.3@osu.edu}
}


%


\maketitle

\begin{abstract}
This brief note highlights a largely overlooked similarity between the SIR ordinary differential equations used for epidemics on the configuration model of a Poisson network and the classical mass-action SIR equations introduced nearly a century ago by Kermack and McKendrick. We demonstrate that the decline pattern in susceptibles is identical for both models. This equivalence carries practical implications: the susceptibles decay curve, often referred to as the epidemic or incidence curve, is frequently used in empirical studies to forecast epidemic dynamics. Although the curves for susceptibles align perfectly, those for infections do differ. Yet, the infection curves tend to converge and become almost indistinguishable in high-degree networks. In summary, our analysis suggests that under many practical scenarios, it's acceptable to use the classical SIR model as a close approximation to the Poisson SIR network model.
\end{abstract}

\begin{IEEEkeywords}
Configuration model;   SIR epidemic equations; Kermack-McKendrick model; 

\end{IEEEkeywords}

\section{Introduction}
In many practical situations, especially during the recent COVID-19 pandemic, it has been suggested that the classical susceptible-infected-recovered (SIR) equations are robust in the sense that  they can accurately depict a wide range of epidemic scenarios, accounting for both heterogeneity of contacts and variations in the infection process. In this brief note, we will consider  one of the surprising examples of this, analyzing an interesting and apparently little-known relation  between the classical SIR models and the network versions of SIR epidemics. As it turns out, in certain situations, the classical SIR equations are also valid when  describing dynamics on specific networks. This fact is somewhat surprising and, at least to some extent, could explain the notable resilience of the classical SIR models, even when they are considered outside  the assumptions they were originally based on or in the areas seemingly far removed from mathematical epidemiology,  like political science and chemical kinetics \cite{volkening2020forecasting,kang2019quasi}. Even though the analysis considered here can be extended more broadly, for simplicity  we specifically focus on an example involving a relatively  simple network known as the configuration model graph with a given degree distribution. For a more comprehensive review of such models in the context of epidemics and more, refer to the monographs \cite{van2016random}, or \cite{kiss2017mathematics}.    In the remainder of this section, we  briefly recall the basic notions relevant to our discussion. The most important results can be found in Section~2, where we establish the equivalence between non-network and network SIR models.
\subsection{Classical SIR Model}
The SIR model, proposed famously by Kermack and McKendrick in \cite{kermack1927contribution}, offers a foundational mathematical approach to understand the dynamics of infectious diseases. Within this model, individuals in a population are grouped into three compartments: susceptible, infected and recovered, with their temporal proportions denoted usually by   \(S(t)\),  \(I(t)\), and  \(R(t)\), respectively. The time progression of these compartments is detailed by the subsequent ordinary differential equations:

\begin{align}\label{eq:sir1}
\dot{S} &= -\beta SI \\ 
\dot{I} &= \beta SI - \gamma I \label{eq:sir2}\\
\dot{R} &= \gamma I. \label{eq:sir3}
\end{align}

Herein, \(\beta\) denotes the rate of infection, representing the frequency at which susceptible individuals become infected upon encountering infected ones. Conversely, \(\gamma\) denotes the recovery rate, defining the pace at which infected individuals recuperate or succumb, subsequently departing from the infected category. The initial conditions for the above system are taken as 
\begin{equation}
    \begin{aligned}
        {S}(0)  &= 1 \\
        {I}(0) &= \rho \\
        {R}(0) &= 0.
    \end{aligned}
    \label{eq:sir_ic}
\end{equation}

A pivotal parameter in epidemiology is the basic reproduction number, \({\cal R}_0 \), defined as:
\begin{equation}\label{eq:r0}  {\cal R}_0 = \frac{\beta}{\gamma}. \end{equation}
This value serves as an indicator of the potential for an outbreak. When \({\cal R}_0 > 1\), it suggests that the epidemic can spread in the population. It's important to note that these equations inherently assume the {\em law of mass action}, without a specific contact structure or, in other words, homogenous interactions among all individuals.  This implies that an individual who is infectious can potentially infect any other susceptible individual. 

\subsection{Network SIR Models}

In real-world populations, interactions among individuals are not homogeneous, a simplifying assumption inherent to classical SIR models. In reality, individuals form complex interaction patterns, often described best as networks. These networks, where nodes represent individuals and edges denote interactions or contacts, embody the diverse and intricate structure of relationships within populations. For instance, some individuals, often called 'super-spreaders', have a disproportionately high number of connections, making them more likely to spread infections. Others may have limited interactions and are consequently less exposed. This  variability in connectivity is  not represent in the system \eqref{eq:sir1}--\eqref{eq:sir3}. By extending the classical SIR model to networks, we can capture these heterogeneities, providing a more accurate representation of disease spread. This network-based perspective has given rise to network versions of the SIR differential equations, offering more nuanced insights into outbreak dynamics in structured populations.  Notable among these models are the pairwise (PW) model~\cite{rand1999correlation,Keeling1999}, the Volz model~\cite{volz2008sir}, and the dynamical survival analysis (DSA) model~\cite{jacobsen2018large,khudabukhsh2020survival}. 

The PW model, as detailed in~\cite{rand1999correlation,Keeling1999}, provides equations that predict the expected number of susceptible nodes and infected nodes, along with the anticipated number of $S-I$  and $S-S$ pairs. This model employs a closure technique that predicts the expected number of triples using singles and pairs, eliminating the need to rely on higher-order moments.

On the other hand, the Volz model~\cite{volz2008sir} relies  on a differential equation system described via  the probability generating function (PGF) of the degree distribution. Unlike other models, it  focuses on edge-centric metrics, like the count of edges connecting nodes in specific states, instead of node-centric metrics such as counts of infected or susceptible nodes. This model's results align closely with simulations. Furthermore, Decreusefond and colleagues  have formally demonstrated that if $N$ is the number of network nodes,  the Volz model represents the large $N$  limit for a stochastic SIR epidemic on a configuration model network~\cite{decreusefond2012large}.

\subsection{DSA Model  and Poisson-type Networks}
In a more recent study, Jacobsen and colleagues  introduced an alternative approach for deriving the mean-field limit of a stochastic SIR model on a  network, which is often referred to  the dynamical survival analysis (DSA) approach~\cite{jacobsen2018large,kiss2023necessary}. Although the variables analyzed  in this approach differ from those in the Volz model, the two limiting network dynamics were shown to be  equivalent \cite{kiss2023necessary}. Moreover, the DSA framework offers a fresh perspective on epidemics by allowing us to see them through a statistical lens, for instance, by estimating the likelihood that a standard node, which was susceptible at time $t = 0$, remains susceptible at any time $t > 0$. 

It turns out that all three models  (PW, Volz and DSA) take a particularly simple and mutually equivalent form when  the network degree  distributions belong to the so-called  Poisson-type class of distributions consisting of the Poisson, negative binomial,  and binomial families of distributions \cite{jacobsen2018large,kiss2023necessary, rempala2023dynamical}. For that class, the large network  equations may be described in terms of  the limiting proportions  of $S$-type nodes ($x_S$), $S-I$ pairs ($x_{SI}$),  and  the infection density $x_D=x_{SI}/x_S$
 \cite{WKB_Hospital23}. 

The essence of the simplification for the Poisson-type networks is that  the limiting equations involve  only  these quantities along with the proportion of infected  $x_I$ as follows :
\begin{align}
        \dot{x}_{S} &= -\wbeta x_{D}x_{S},  \nonumber\\ 
        \dot{x}_{D} &= \wbeta (1 - \kappa) x_{D}^2 + \left(\wbeta \kappa \mu x_{S}^{2\ka - 1} - (\wbeta + \wgamma)\right) x_{D} \label{eq:nsir}\\
           \dot{x}_{I} &= \wbeta x_{D} x_{S} - \wgamma x_{I},\nonumber  
    \end{align}
 with the  set of initial conditions 
\begin{equation}
    \begin{aligned}
        x_{S}(0)   &= 1 \\
        x_{I}(0) &= \wrho \\
        x_{D}(0) &= \mu \wrho.
    \end{aligned}
    \label{eq:nsir_ic}
\end{equation} For simplicity, we omit here the dynamics of recovery, which is the same as in the classical model \eqref{eq:sir3}.  
Note that the constants $\wbeta$ and $\wgamma$  represent the  rates of infection and recovery,  with $\wrho$ representing the initial amount of infection.  The quantities  $\mu$ and $\ka$ are network-related and represent, respectively,  the average of the network degree  and  the ratio of the average degree to the average {\em excess degree}. Recall that the  excess degree distribution $q_k$ on a configuration model random graph with degree distribution $p_k$  and average degree $\mu<\infty$  is defined as the degree of a random neighbor node of a randomly selected node of degree at least one. That is, $q_k=(k+1) p_{k+1}/\mu$. For the network SIR model \eqref{eq:nsir},  the basic  reproduction number \eqref{eq:r0} needs to be redefined to reflect the added heterogeneity.    To avoid ambiguity, we denote this basic reproduction number by \(\Rnet \) and  the basic reproduction number given in  \eqref{eq:r0} by \(\Rma\). The network basic reproduction number is then 
\begin{equation}\label{eq:r0net}
\Rnet=\frac{\mu \wbeta}{\wbeta+\wgamma}
\end{equation} and reflects the somewhat  intuitive notion  that the average number of new infections has to be proportional to the average of the  degree distribiution.

\section{Models Equivalence}
Let us consider the case of a Poisson network, that is, the special case when $p_k=\exp(-\mu) \mu^k/k! $ for $k\ge 0$ are Poisson probabilities. In this case it is easy to see that we have $q_k=p_k$ and thus $\ka=1$. Consequently, the network equations \eqref{eq:nsir} simplify to 
\begin{align}
        \dot{x}_{S} &= -\wbeta x_{D}x_{S},  \label{eq:nsir1}\\
          \dot{x}_{D} &=  \wbeta  \mu x_Dx_S - (\wbeta + \wgamma) x_{D} \label{eq:nsir2}\\
        \dot{x}_{I} &= \wbeta x_{D} x_{S} - \wgamma x_{I}, \label{eq:nsir3}    
    \end{align} with the same set of  initial conditions \eqref{eq:nsir_ic}.
These   equations may be manipulated  in order to yield only a single equation describing the dynamics of susceptibles known as the  {\em epidemic curve equation}. 
\subsection{Epidemic Curve Equations}
\begin{figure}[htbp] 
   \centering
   \includegraphics[width=3.5in]{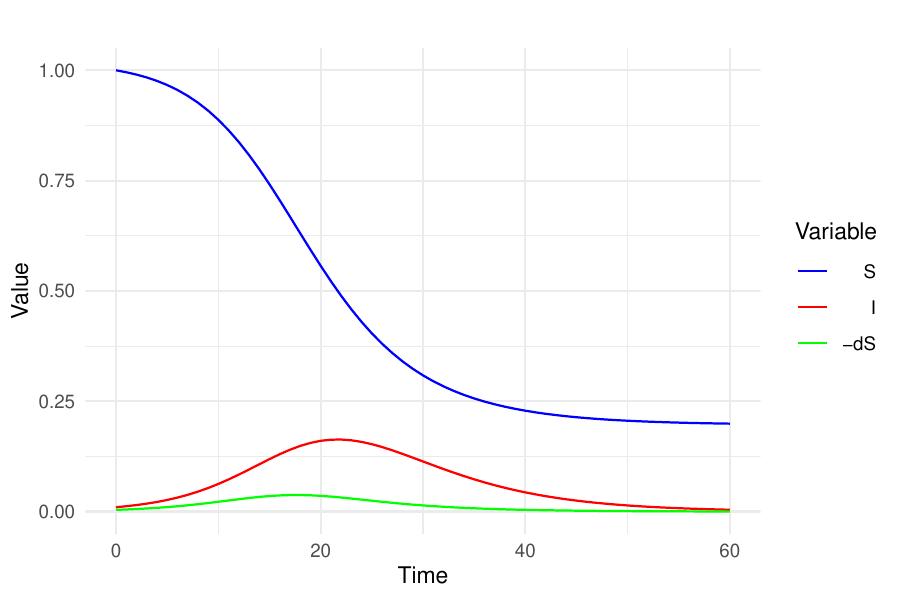} 
   \caption{{\bf SIR Curves.} SIR model curves from \eqref{eq:sir1}-\eqref{eq:sir3} for the set of parameters $\beta=0.4$, $\gamma=0.2$, and $\rho=0.01$. The lowest curve is the epidemic curve $-\dot{S}$ given in \eqref{eq:ma_ec}. }
   \label{fig:1}
\end{figure}

Dividing  \eqref{eq:nsir2} by \eqref{eq:nsir1} we obtain 
\begin{equation*}
    \frac{d{x}_{D}}{d{x}_{S}} 
    = - \mu + \frac{\mu}{\Rnet} \frac{1}{x_{S}},
\end{equation*}
and therefore (applying the initial conditions \eqref{eq:nsir_ic}) we get 
\begin{equation}\label{eq:den}
x_{D} = \mu \left(1+\wrho -x_S-\ln x_S/\Rnet\right).
\end{equation}
Substituting this expression into \eqref{eq:nsir1} we obtain the following differential equation which involves 
only $x_S$ 
\begin{equation}
\label{eq:n_ec}
-\dot{x}_S=\mu\wbeta \left(1+\wrho -x_S-\ln x_S/\Rnet\right)x_S.
\end{equation}
The equation above inherits the initial condition 
$x_S(0)=1$ from \eqref{eq:nsir} and  is seen as describing the rate of decay of the susceptible population in terms of the number of susceptibles only. It is especially relevant to analyzing new infections or the {\em incidence rate} of an epidemic. The quantity $-\dot{x}_S$ is therefore often referred to as  the {\em epidemic curve}.  If we think about the quantity $x_S(t)$ as the survival probability for susceptibles (that is, the probability of a randomly selected initially susceptible avoiding infection by time $t>0$, see, e.g.,    \cite{kiss2023necessary}) then the epidemic curve may be also viewed as the {\em density of infection}, see, e.g., \cite{rempala2023dynamical}.  An example of the survival curve and the corresponding density of infection (or epidemic curve) is presented in Figure~\ref{fig:1}. 

The equation \eqref{eq:n_ec} can be also used as a convenient way of comparing the dynamics of the Poisson network SIR model \eqref{eq:nsir1}-\eqref{eq:nsir3} with that of the classical (mass action) SIR model \eqref{eq:sir1}-\eqref{eq:sir3}. Indeed,  note that since the algebraic structures of the right hand sides of  the differentrial equations \eqref{eq:nsir1}-\eqref{eq:nsir2} and  \eqref{eq:sir1}-\eqref{eq:sir2} are identical,  the same manipulation  as for the former may also be used for  the latter, leading to  the mass action epidemic curve equation 
 \begin{equation}
\label{eq:ma_ec}
-\dot{S}=\beta \left(1+\rho -S-\ln x_S/\Rma\right)S.
\end{equation}
The equations \eqref{eq:ma_ec} and \eqref{eq:n_ec} both describe the epidemic curve, they use however different assumptions on the population contact structure, namely the mass action and  Poisson degree distribution, respectively.  The two equations are seen to be  equivalent if and only if   the appropriate parameter values coincide. Note also that in this case $S=x_S$ but  $I\ne x_I$.  Let us formulate this as follows. 
\begin{thm} Assume that we wish to approximate the dynamics of a Poisson network SIR epidemic  given by \eqref{eq:nsir1}-\eqref{eq:nsir3} and \eqref{eq:nsir_ic} using the classical SIR mass action model \eqref{eq:sir1}-\eqref{eq:sir3} and \eqref{eq:sir_ic}. The two models respective  epidemic curve equations coincide iff $\beta=\mu\wbeta$, $\rho=\wrho$,  and $\gamma=\wgamma+\wbeta$.  In this case $S=x_S$,  $I=x_D/\mu$, and 
$\Rma=\Rnet=\R$.
Moreover, if $\R<\mu$  the true amount of infection under network model  $(x_I)$ satisfies 
\begin{equation}\label{eq:cor}\dot{x}_I=-\dot{x}_S-\gamma\theta x_I \end{equation} where the correction factor $\theta$ is given by 
 $$ \theta=1-\R/\mu.$$ 
 \qed
\end{thm}
The result above indicates that it is reasonable to analyze  the epidemic curve on the Poisson network using the classical SIR equations \eqref{eq:sir1}-\eqref{eq:sir3}. This will lead to the correct dynamics of the incidence (infection density) given by $x_S$ curve. However, in order to correctly calculate the number of infections $x_I$, the classical equation \eqref{eq:sir2} needs to be replaced by the corrected   one as given in \eqref{eq:cor}.
The discrepancy will depend on the values of the parameters but may be considerable as illustrated in Figure~2. However, for large $\mu$ we see that $\theta\approx 1$ and the discrepancy becomes  negligible  in practice. 

\begin{figure}[htbp] 
   \centering
   \includegraphics[width=3.5in]{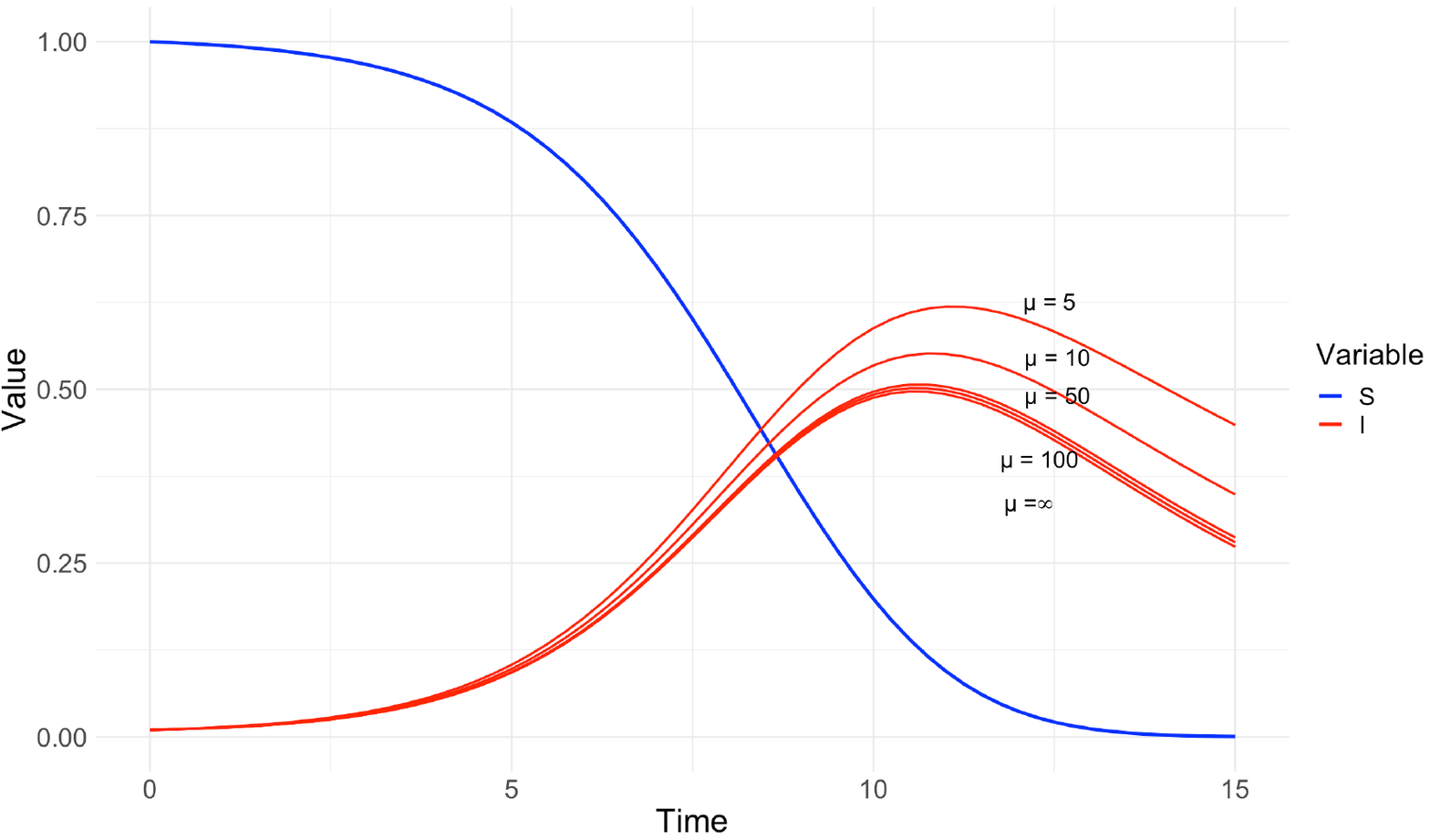} 
   \caption{{\bf Approximating Network Model.} Poisson network SIR model infected curves (red) from \eqref{eq:nsir1}-\eqref{eq:nsir3}  plotted for  several different mean degrees ($\mu$).  The remaining  parameter values are as in  Figure~1, with the network parameters matched as in Theorem~1.  It is seen that for large  $\mu$ (at least 50) the  network   curves  get close  to the one classical SIR model one, that is when $I=x_I$ and  correction factor $\theta=1$ ($\mu=\infty$). At teh same time according to Theorem~1 the susceptible curve (blue) is the same  for  both models ($S=x_S$). }
   \label{fig:2}
\end{figure}

\section{Conclusions}

Network-based epidemic models often employ random graphs with Poisson degree distributions. This choice is grounded in the widely applied Poisson approximation of the binomial distribution, as applied to  the basic Erd{\"o}s-R{\'e}nyi graph model. Personal network analysis based on such random graphs has gained prominence in modeling disease spread, a relevance accentuated by the recent COVID-19 pandemic. In this note, we pointed out to the surprising equivalence between the classical Kermack and McKendrick SIR model and its Poisson network counterpart. Our result stated in Theorem~1  indicates that SIR epidemics for Poisson networks, particularly those with a high degree, can be effectively represented using the traditional mass-action SIR dynamics. This compatibility is evident in epidemic curves and infection magnitude analyses as illustrated in our Theorem~1.  The observed congruence between the classical and network-based SIR models offers a valuable perspective for models simplification especially for  the  high-degree Poisson networks that became especially relevant in assessing proximity-based transmission during COVID-19 pandemic. While the  insights presented here  are directly relevant to epidemiology, they also have implications in other fields, including social networks analysis and modeling dynamics of  experimental systems in chemical physics.

\bibliographystyle{IEEEtran}
\providecommand{\noopsort}[1]{}

\end{document}